\documentstyle[12pt,aasms4]{article}

\def\arcsecpoint{$''\!.$}

\received{1998 February 18}
\accepted{1998 May 16}

\slugcomment{To appear in {\it The Astrophysical Journal}}

\lefthead{Kraemer, Ruiz, \& Crenshaw}
\righthead{Narrow-Line Region of NGC 1068}

\begin{document}

\title{Physical Conditions in the Inner Narrow-Line Region\\
of the Seyfert 2 Galaxy NGC 1068\altaffilmark{1}}

\author{Steven B. Kraemer\altaffilmark{2,3},
Jos\'{e} R. Ruiz\altaffilmark{2},
\& D. Michael Crenshaw\altaffilmark{2}}

\altaffiltext{1}{Based on observations made with the NASA/ESA Hubble Space 
Telescope, obtained from the data archive at the Space Telescope Science 
Institute. STScI is operated by the Association of Universities for Research in 
Astronomy, Inc. under the NASA contract NAS5-26555. }

\altaffiltext{2}{Catholic University of America,
NASA/Goddard Space Flight Center, Code 681,
Greenbelt, MD  20771.}

\altaffiltext{3}{Email: stiskraemer@yancey.gsfc.nasa.gov.}

\begin{abstract}

  The physical conditions in the inner narrow line region (NLR) of the Seyfert
  2 galaxy, NGC 1068, are examined using ultraviolet and optical spectra 
  and photoionization models. The spectra are {\it Hubble Space Telescope} ({\it HST})
  archive data obtained with the Faint Object Spectrograph (FOS). We selected 
  spectra of four regions, taken through the 0\arcsecpoint3 FOS aperture, covering the 
  full FOS 1200 \AA\ to 6800 \AA\ waveband. Each region is approximately 20 pc 
  in extent, and all are within 100 pc of the apparent nucleus of 
  NGC 1068. The spectra show similar emission-line ratios from wide range 
  of ionization states for the most abundant elements. After extensive 
  photoionization modeling, we interpret this result as an indication that 
  each region includes a range of gas densities, which we included in the models as separate 
  components. Supersolar abundances were required for several elements to fit 
  the observed emission line ratios. Dust was included in the models but 
  apparently dust to gas fraction varies within these regions. The low 
  ionization lines in these spectra can be best explained as arising in gas 
  that is partially shielded from the ionizing continuum. 

  Although the predicted line ratios from the photoionization models provide a 
  good fit to the observed ratios, it is apparent that the model predictions of 
  electron temperatures in the ionized gas are too low. We interpret this as 
  an indication of additional collisional heating due to shocks and/or 
  energetic particles associated with the radio jet that traverses the NLR of 
  NGC 1068. The density structure within each region may also be the 
  result of compression by the jet.

\end{abstract}

\keywords{galaxies: individual (NGC 1068) -- galaxies: Seyfert}

\section{Introduction}

   NGC 1068, one of the initial set of emission line galaxies studied
   by Seyfert (1943), is the nearest (z=0.0036) and the best studied 
   of the Seyfert 2 galaxies. NGC 1068 has been observed extensively in all
   wavebands from the radio to the X-ray. Not only is there evidence of 
   ionizing radiation from the central AGN, but a prominant radio jet exists in
   the NLR (Wilson \& Ulvestad 1983), and there is a starburst ring 
   approximately 1 kiloparsec from the nucleus (Snijders, Briggs, \& Boksenberg 
   1982, Bruhweiler, Troung, \& Altner 1991). The detection of polarized optical 
   continuum and broad permitted lines (Miller \& Antonucci 1983, Antonucci \& 
   Miller 1985) in the nucleus of NGC 1068 was the inspiration for the 
   "unified model" for Seyfert galaxies, in which the differences between
   types 1 and 2 were attributed to viewing angle (Antonucci 
   1994), with Seyfert 2's characterized by obscuration of their central 
   engines. 

   Due to its relative proximity, NGC 1068 offers unique opportunity
   to study the detailed physics of the NLR gas. Studies of the
   conditions in the NLR can provide a check on the estimates of the
   luminosity and spectral characteristics of the intrinsic ionizing
   continuum proposed by Pier et al. (1994) and Miller, Goodrich, \& Mathews (1991). Ground based 
   observations and analysis of the extended narrow line region (Balick \& Heckman 1985, 
   Evans \& Dopita 1986, Bergeron, Petitjean, \& Durret 1989) have shown
   that gas at large distances from the nucleus ($\sim$ kiloparsecs) is likely 
   to be photoionized by the radiation from the central active galactic nucleus
   (AGN). It follows that photoionization must an important if not dominant 
   process in the inner NLR of NGC 1068 as well. Before {\it HST}, it was difficult to 
   examine the conditions in the inner 100 pc. Ground based optical 
   spectra (cf. Koski 1978) and UV spectra from {\it IUE} (Snijders et al. 1982) that 
   sampled large areas ($\geq$ 3") within NGC 1068 did not have the resolution to
   provide the necessary constraints on models of the emission line gas.  With 
   {\it HST} resolution and spectral coverage, we now have access to
   spatially-resolved spectra of the inner NLR. 

   Although it is clear that the NLR gas is photoionized (Netzer 1997), 
   it is possible that collisional processes are important as well (Kriss et al. 
   1992). Detailed photoionization models can help distinguish between the 
   contributions of various possible sources of ionization and heating in the 
   NLR. A better determination of the relative contributions of such process 
   may lead to an understanding of the physical nature of the NLR, and possibly 
   its origin and evolution. Even though the ionizing continuum in 
   NGC 1068 cannot be directly observed, the relative radial distances and 
   physical extents of the regions observed are known, important new 
   constraints on the models.

\section{Observations and Analysis}

There are a large number of FOS observations of NGC 1068 in the {\it HST} archives. 
FOS spectra were obtained of the brightest point in the visible region, which we 
refer to as the ``nucleus''; some of the spectra have been published by
Caganoff et al. (1991) and Antonucci et al. (1994).Spectra were also obtained 
at various offset positions from the nucleus, but have not been published.
In the {\it HST} archives, these positions are often referred to as ``clouds''
(e.g., ``Cloud 1''), but we will refer to them as ``positions'' (e.g., 
``Position 1''), since FOC and WFPC2 [O III] images show that even the FOS
0\arcsecpoint3 aperture encompasses a number of emission-line knots.
We chose to limit the number of spectra for this study to satisfy several 
criteria. First, the position observed must have full wavelength coverage from  
1200 -- 6800 \AA\ at good resolution ($\lambda$/$\Delta$$\lambda$ $\approx$ 
1000), to provide a full range of emission-line diagnostics; thus only pointings 
that include observations with the G130H, G190H, G270H, G400H, and G570H 
gratings were used. Second, we wanted to concentrate on regions of small spatial 
extent, to minimize the range in physical conditions, so only observations 
through the 0\arcsecpoint3 aperture were included. Finally, for observations 
obtained of the same region at different times, the acquisition techniques had 
to be the same, and we required that the spectra match well where they 
overlap. Table 1 gives a summary of the observations that we used. Note that 
observations of ``NGC1068'' and ``NGC1068-NUC'' are of the same region (the 
``nucleus'').

Our requirements resulted in UV and optical spectra of the nucleus and 
three offset positions in the inner NLR through a 0\arcsecpoint3 aperture. The 
nucleus was acquired by peakups through successively smaller apertures 
(1\arcsecpoint0, 0\arcsecpoint5, and 0\arcsecpoint3) using light from the G270H 
grating. Spectra of the other positions were obtained by offsetting to positions 
of bright emission seen in the original WF/PC narrow-band images (centered on 
various emission lines such as [O III]). Peakups and spectra were obtained with 
the FOS/BLUE detector and G130H and G190H gratings, and with the FOS/RED 
detector and G270H, G400H, and G570H gratings. In addition, blue G270H spectra 
were obtained of the nucleus after peakups with the blue detector. In this case, 
after scaling the red G270H spectrum by a factor of 1.1, the features in both 
blue and red G270H spectra are essentially identical. Thus, we are confident 
that peakups on the nucleus performed at different times and with different 
detectors resulted in observations of the same region. Since the offsets are 
highly accurate, and the spectra at each offset position matched to 
within 10\% in the wavelength regions of overlap, we are confident of these 
pointings as well. To account for the small ($\leq$ 10\%) absolute flux 
differences, all of the spectra from a given position were scaled to match the 
flux level of the red G270H spectra. 

Figure 1 shows the locations of the aperture for these pointings, superimposed 
on an FOC [O III]$\lambda$5007 image obtained from the {\it HST} archives and 
the axis of the radio jet (Gallimore et al. 1996).
We note that the FOS observations were obtained prior to the 
installation of COSTAR on {\it HST} in 1993 December, so that a substantial amount of 
light from outside the projected aperture is included in each of these spectra. 
These effects are discussed in greater detail below. At the distance of NGC 
1068, 0\arcsecpoint3 corresponds to 21 pc (for z=0.0036 and H$_{0}$ = 75 km 
s$^{-1}$ Mpc$^{-1}$).

Figure 2, 3, and 4 show the far-UV, near-UV, and optical spectra for each 
position. We note that the N V $\lambda\lambda$1239, 1243 and C IV 
$\lambda\lambda$ 1548, 1551 doublets, normally blended together in 
Seyfert 2 spectra, are resolved in some of the far-UV spectra, since we are 
isolating specific kinematic regions and these lines are therefore relatively
narrow. The blue component of the doublet {\it appears} to be smaller than the 
red component in each case, because Galactic and intrinsic absorption features 
(probably from the halo of NGC 1068) are absorbing 
the blue side of the doublet emission in each case. We also note that the 
spectra from different positions are very similar in appearance, given the 
differences in absolute fluxes, and that the N V and [Ne V] $\lambda\lambda$3346, 
3424 lines are unusually strong in each spectrum. We will explore these issues 
later in the paper. 

We measured the fluxes of most of the narrow emission lines by direct 
integration over a local baseline determined by linear interpolation between 
adjacent continuum regions. For severely blended lines such as H$\alpha$ and [N~II] 
$\lambda\lambda$6548, 6584, we used the [O~III] $\lambda$5007 profile as a 
template to deblend the lines (see Crenshaw \& Peterson 1986).
We then determined the reddening of the narrow emission lines from the observed 
He~II $\lambda$1640/$\lambda$4686 ratio, the Galactic reddening curve of Savage 
\& Mathis (1979), and an intrinsic He II ratio of 7.2, which is expected from 
recombination (Seaton 1978) at the temperatures and densities typical of the NLR 
(see also Section 3).
We determined errors in the dereddened ratios from the sum in quadrature of the 
errors from three sources: photon noise, different reasonable continuum 
placements, and reddening. 

Table 2 gives the dereddened narrow-line ratios, relative to 
H$\beta$, and errors in the dereddened ratios for each position. Inspection of 
this table shows that the emission-line ratios from the different regions
are indeed very similar. At the end of the table, we give the H$\beta$ fluxes 
(ergs s$^{-1}$ cm$^{-2}$) in the aperture and the reddening values that we 
determined from the He II ratios.

As we mentioned earlier, these observations were obtained prior 
to the installation of COSTAR . Hence, the presence of broad wings on the 
point-spread function at the aperture plane leads to substantial 
contamination of the observed flux by emission-line knots outside of the 
projected aperture. To estimate this effect on the observed spectra, we 
retrieved (from the STScI) a model pre-COSTAR point spread function (PSF) for 
the FOS red detector at 5000 \AA, which was generated and described by Evans 
(1993). We interpolated over the reseaux in the post-COSTAR FOC [O III] image 
in Figure 1 (which has a spatial resolution of 0\arcsecpoint014 per pixel), 
extracted subimages through apertures of different sizes, and convolved the 
original image and subimages with the FOS PSF image. We then determined the 
percentage of the [O III] flux in the spectrum of each region that is 
contributed by emission within the projected 0\arcsecpoint3 aperture, within a concentric 
aperture of diameter 0\arcsecpoint6, and from the remainder of the NLR flux in 
the FOC image. These values are respectively 61\%, 22\%, and 17\% for the 
nucleus; 53\%, 30\%, and 17\% for position 1; 54\%, 25\%, and 21\% for 
position 2; and 65\%, 20\% and 15\% for position 3. Thus, in these cases, we are sampling regions considerably 
larger than the projected aperture sizes; however, in each pointing, at least $\sim$80\% of the 
[O~III] flux is coming from within $\sim$0\arcsecpoint6 of the aperture centers. 
Position 1 is $\sim$0\arcsecpoint2 from the ``nucleus'', so we are primarily 
sampling the same region in this case. For positions 2 and 3, we are sampling 
regions that are relatively distinct, since they are 0\arcsecpoint6 and 
1\arcsecpoint5 from the nucleus. 

\section{Photoionization Models}

   As in our previous studies (e.g., Kraemer et al. 1998), we have taken
   a simple approach in setting the initial input values for photoionization
   models. In previous cases where we could not resolve the emission-line 
   region, we could effectively adjust the distance of the emission-line gas 
   from the source of the ionizing radiation to obtain a good match with
   the obervations. With the current
   data set we have spectra from spatially resolved regions and, therefore, 
   the distance of the gas from the ionizing source is more tightly constrained.
   In turn, we can be more flexible in adjusting input parameters such as 
   abundances and optical thickness, if our initial parameters are not
   sufficient. Such an approach can produce a better fit to the observed line 
   ratios, but it is important to bear in mind that it still might not result 
   in a unique solution for the set of physical conditions in the emission 
   line region. We will return to this point in the following section.
   
   The photoionization code used for this study has been described in
   detail in our previous papers (cf. Kraemer et al. 1998), and we will not
   repeat the description here. A few points should be mentioned, however.
   First, it is important to bear in mind that this code assumes a slab
   geometry, with photon escape out the illuminated face. The effects of
   dust are included, including internal reddening, trapping of UV resonance
   lines and screening of the ionization radiation. Forward scattering
   by the grains is assumed. For a full explanation of the treatment of
   dust in this code, see Kraemer (1985). As per the standard convention,
   models are parameterized in terms of the density of atomic hydrogen
   (N$_{H}$) and the dimensionless ionization parameter at the illuminated face
   of the cloud:
 
\begin{equation}
U = \int^{\infty}_{\nu_0} ~\frac{L_\nu}{h\nu}~d\nu ~/~ 4\pi~D^2~N_{H}~c,
\end{equation}

   where L$_{\nu}$ is the frequency dependent luminosity of the ionizing continuum,
   D is the distance between the cloud and the ionizing source and 
   h$\nu$ =13.6 eV. 
     
\section{Choosing the Model Input Parameters}

   In spite of the pre-COSTAR PSF problems described in section 2, we know 
   that the majority of the line emission seen in these spectra arise in or 
   near the projected aperture locations shown in Figure 1. The spectra from
   each of these regions show a wide range of ionization states for the most 
   abundant elements, indicating gas with a range of physical conditions 
   within each region. In simple multi-component models of an entire NLR 
   (see Kraemer et al. 1994), one can include contributions from gas in various 
   ionization states by placing the components at different distances from the ionizing
   source. This means that there can be a large amount of low-ionization, low density
   gas at large radial distances; having a large volume of this gas 
   can balance the fact that such components have low emissivity relative to
   higher ionization denser material close to the ionizing source, as long
   as the covering factor of the inner component is small. We do not have
   this flexibility in modeling these spectra. If the line emission is from 
   photoionized gas, there must be a range in density within the observed
   regions. After running an initial set of simple multicomponent models, it
   became clear that we needed additional parameters to match the observations.
   These parameters include elemental abundances, "shielding" of one of
   our components, optical depth, and dust.

   Although previous photoionization studies have suggested that the elemental
   abundances in the NLR may be non-solar (Osterbrock 1989), line ratios
   indicative of non-solar abundances can often be explained by including
   multi-component models of varying density. For these data, however,
   there are clear indications of non-solar abundances. For example,
   as discussed by Netzer (1997), the ratio of the O~III] $\lambda$1664/
   N~III] $\lambda$1750 lines
   can be used to estimate the ratio of elemental oxygen to nitrogen, since
   their ionization regions tend to show large overlap. The theoretical
   ratio is as follows:

\begin{equation}
\frac{I(\lambda1664)}{I(\lambda1750)} = 0.41T_{4}^{-0.04}exp(-0.43/T_{4}) \frac{N(O^{+})}{N(N^{+})}
\end{equation}

   where T$_{4}$ is the temperature in units of 10,000K. The average intensity
   ratio  from the four regions sampled is 0.65.
   If we assume T=15,000K, N(O)/N(N) is $\sim$2.1, which is 0.36 times solar. Netzer
   interprets this as a large oxygen underabundance, and offers the
   observed O~III] $\lambda$1664/C~III] $\lambda$1909 ratio as further evidence, since it
   yields an N(O)/N(C) $\sim$ 0.73. Averaging over the four regions, we
   obtain N(O)/N(C) $\sim$ 1.8, which is approximately solar. Therefore, 
   it is likely that we are seeing enhanced nitrogen, rather than depleted 
   oxygen.

   Another indication that the nitrogen is supersolar is the ratio of the
   N~V$ \lambda$1240 line to the He~II $\lambda$1640. In photoionized gas, the 
   N~V/He~II ratio
   is typically less than unity (Ferland et al. 1996). Although this ratio can increase
   if the gas is so optically thin that the edge of the He$^{++}$ Stromgren zone
   is never attained, in such cases the O~V $\lambda$1216 and O~VI $\lambda$1035 lines become
   inordinately strong (Netzer was able to achieve such ratios without excess
   O~V and O~VI emission by depleting oxygen by a factor of 3, which we do not 
   think is supported by the observations). Ferland et al. suggest that 
   relative enhancement of nitrogen can better explain such line ratios. 
   Although their study was of the spectra of luminous QSOs, not only are the 
   same physical processes at work in the ionized gas, it is not too surprising 
   that there may be heavily reprocessed material in the nucleus of a Seyfert
   galaxy, since it is possible that the AGN phenomenon was proceeded by a 
   massive nuclear starburst (cf. Osterbrock 1993) Finally, the [N~II] 
   $\lambda\lambda$ 6548,6584 lines are quite strong in these spectra (4 to 5 
   times the strength of H$\beta$) while the [O~II] $\lambda$3727
   line is weak. Although this is possibly due to collisional
   suppression of the [O~II] line, it is also plausible that we are seeing
   the effect of overabundance of nitrogen. This will be explained in more
   detail in the discussion of the model results. For the models, we
   assume a three-times solar nitrogen abundance.

   The lines of Ne$^{+3}$, Ne$^{+4}$ and Fe$^{+6}$ are quite strong in all these spectra.
   Fitting the coronal lines with solar abundances is often a problem
   for photoionization models (Kraemer et al. 1998), but as Oliva (1997)
   suggests, they can be enhanced if these elements are overabundant. 
   Furthermore, from their analysis of {\it ASCA} data, Netzer \& Turner (1997) 
   postulate that the Fe/O ratio is quite high in the X-ray emitting gas in 
   NGC 1068. For these models, we have assumed that both iron and neon are 
   supersolar by a factor of 2 in abundance.

   Although it is possible that other elements may be overabundant, 
   there is no indication from the spectra that this is the case. Therefore,
   we have chosen to keep them at solar abundances. The numerical abundances,
   relative to hydrogen, assumed for these model are as follows: He=0.1,
   C=3.4x10$^{-4}$, O=6.8x10$^{-4}$, N=3.6x10$^{-4}$, Ne=2.2x10$^{-4}$, 
   S=1.5x10$^{-5}$, Si=3.1x10$^{-5}$, Mg=3.3x10$^{-5}$, Fe=8.0x10$^{-5}$.

   We have assumed in these simple models that the gas is photoionized
   by radiation from the central AGN. In NGC 1068, as in most Seyfert 2
   galaxies, it is impossible to measure the intrinsic ionizing continuum
   directly, since the inner regions of these objects are usually obscured
   by a large column of dusty gas. The ionizing continuum is only observed
   by light scattered into our line of sight by a scattering medium, 
   possibly consisting of free electrons (see Antonucci 1994 for the details
   of this basic model). There have been attempts (Miller et al. 1991, Pier et
   al. 1994) to determine the intrinsic luminosity and spectral energy 
   distribution (SED) of NGC 1068 based on observations in the non-ionizing UV 
   and  X-ray, assumptions 
   about the nature of the scattering medium, and comparison to AGN whose 
   nuclei can be observed more directly. The results of these two papers are 
   similar; we have chosen an SED similar to that assumed by Pier et al. as it 
   is the simpler of the two. It consists of a broken power, F$_{\nu}$$=$K$\nu^{-\alpha}$, 
   where:

\begin{equation}
    \alpha = 1.6,~ 13.6eV \leq h\nu < 2000eV
\end{equation}
\begin{equation}
    \alpha = 0.5, ~h\nu \geq 2000eV
\end{equation}

   In addition, we have taken the observed fluxes at log($\nu$) = 15.376 and 
   17.684 quoted by the authors and assumed the same value for the fraction 
   of intrinsic light reflected into our line of sight, f$_{rel}$=0.015. Integrating over frequency 
   and dividing by f$_{rel}$ yields a luminosity in ionizing photons of 
   4x10$^{54}$ sec$^{-1}$, which is typical of Seyfert 1 nuclei.

   As mentioned above, the range of emission lines seen in the spectra
   of each of these regions indicate a mix of physical conditions. Although
   emission from a wide range of ionization states is possible from a
   single component characterized by one atomic density, the most highly
   ionized parts of such a region would have the highest emissivity (greatest
   electron density and temperature) and therefore would dominate the 
   integrated spectrum. Since we see strong lines from both low and high
   ionization states, it is likely that the physical properties of the
   regions where these lines form are indeed different, and that we are
   seeing emission from distinct regions.

   An initial guess at temperature and density can be derived from the 
   ratio of [O~III] $\lambda\lambda$5007,4959/[O~III] $\lambda$4363 (Osterbrock 1974). Averaged
   over the four spectra, this ratio is $\sim$ 46, indicating a temperature
   in excess of 20,000K in the low density limit. It is difficult
   to obtain such a high electron temperature in the O$^{++}$ zone in photoionized
   gas. It is possible that we are seeing a modest modification of 
   this ratio by collisional effects. If we assume a density of 1x10$^{5}$cm$^{-3}$
   , the observed [O~III] line ratio yields a temperature = 17,000K, which is
   more characteristic of the O$^{++}$ in photoionized gas. Thefererefore, we assign
   this density to one component in these models. It is important, however, 
   to note that this simple assumption may not be correct if mechanisms other 
   than photoionization contribute to the thermal balance in the emission
   line gas, as we shall discuss in section 7.
   
   The gas in which the bulk of the [N~II] emission is formed must be 
   characterized by a lower ionization parameter. There are 
   three ways in which U can be lowered: increase D, increase N$_{H}$, or 
   decrease L. 
   Since the individual regions are small ($\sim$20pc) and a 
   few tens of parsecs away from the putative central source, there cannot be 
   sufficient range in distance to account for this drop. Since 
   the O$^{++}$ region is best characterized by a density of 10$^{5}$cm$^{-3}$,
   increasing the density
   of the lower excitation gas would weaken the [N~II] emission 
   due to collisional de-excitation. The simplest
   explanation is that the low excitation gas must see a different ionizing
   continuum. Although one possibility is local sources of ionizing radiation
   (Axon et al. 1998), we propose an alternative. 
   The low ioinzation (N$^{+}$) gas is screened from the central source by the 
   O$^{++}$ gas,
   it is therefore ionized by a filtered continuum. Gas of the same, or lower,
   density, will then be in a lower state of ionization. Ferland \& Mushotzky
   (1982) proposed that the NLR of NGC 4151 is illuminated by radiation
   that is partly absorbed by the BLR gas (the so-called "leaky absorber"
   model). The leaky absorber SED is much harder than the intrinsic SED
   of the galaxy, and the conditions in the NLR gas are strongly influenced
   by the affects of X-ray ionization. Collisional excitation of Ly$\alpha$
   and H$\alpha$ become important processes in such gas. Also, extended partially
   ionized zones can form. Evidence for both these affects can be seen
   in this set of spectra. The larger than Case B H$\alpha$/H$\beta$ ratio is most 
   likely due to collisonal enchancement of H$\alpha$. Also, if the emissivity of 
   the gas in which the [NII] lines form is low compared to that in which the
   higher excitation lines form, there must be a large volume of it to produce
   low excitation lines of comparable strength. We found in generating these
   models, that placing gas of lower density behind the O$^{++}$ region was more
   likely to produce extended zones of N(H$^{+}$)/N(H) $\approx$ 25\%. At very low
   density the emissivity of the gas was so low that the emitting regions
   had to be much larger than the sizes of the regions observed. Note that,
   in this type of model, the covering factor of the low excitation gas
   must be the same as the high excitation gas, so the area of this component
   is constrained. We found that this component could be best modeled by 
   assuming a density of 5x10$^{4}$cm$^{-3}$ and an input spectrum filtered 
   through a column density of $\sim$ 2x10$^{21}$cm$^{-2}$. 

   Finally, there must be another component in which the highest excitation
   lines, in particular N~V $\lambda$1240, arise. For the sake of simplicity, we assigned
   the same density to this component as used for the low excitation gas
   (5x10$^{4}$cm$^{-3}$). Although this is arbitrary, it does produce a simple and
   self-consistent model. Specifically, there is a component of gas of
   density 5x10$^{4}$cm$^{-2}$, which is partially shielded from the central
   source by a higher density component. The high density gas is relatively
   optically thick, but is physically thin ($\sim$ 10$^{-2}$ parsecs). The covering
   factors of the high density component and the shielded gas are equal. 
   The relative covering factors of the high density gas and the unshielded
   lower density gas may vary among the regions.

   In each of the four spectra, the ratio Ly$\alpha$/H$\beta$ is less than 
   20. The low density ratio from recombination is 24. Therefore, it is likely
   that dust mixed in with the emission line gas is responsible
   for the destruction of the Lyman alpha photons. The strength of the other
   resonance lines, specifically N~V $\lambda$1240 and C~IV $\lambda$1550, indicate that there
   cannot be much dust in the most highly ionized gas. Furthermore,
   the presence of the Fe$^{+6}$ and Mg$^{+}$ lines infer that the depletion of these
   refractory elements into grains cannot be near total, as it is in the
   Galactic interstellar medium (Seab \& Shull 1983). Therefore, we have 
   assumed different dust fractions in each of the three components in this 
   model, as suggested by Netzer (1997). The highest ionization gas is dust free. The low ionization gas 
   has a fraction of graphite dust 30\% that found in the galactic interstellar 
   medium. The medium ionization component is quite dusty with dust fraction of 
   graphite and silicate dust 75\% and 50\% the galactic value, respectively. 
   The depletions of carbon, oxygen, silicon, magnesium, and iron are
   scaled by these dust fractions, assuming galactic interstellar medium 
   values: 80\% for carbon, 20\% for oygen and complete depletion for the 
   refractory elements. The dust fractions assumed are arbitrary, but the 
   model results are not particularly sensitive to the exact fraction of the 
   different types of dust. The main point for these models is that assuming a 
   mix of dusty and dust-free gas yields the best fit to the observed line
   ratios.

\section{Model Results}

   Our approach in modeling these regions in NGC 1068 was to to fit 
   the O$^{++}$ emission gas first, and then add the low excitation component. This was,
   of course, necessitated by the fact that we have assumed the low excitation
   gas is ionized by a continuum filtered by the O$^{++}$ component. Once these
   two models were complete, we added the third component, primarily to
   fit the N~V $\lambda$1240 line. Having arrived at the densities of these three
   components as described above, we set the ionization parameters for the
   models. In order to avoid adding additional components, we set the
   ionization parameter high enough for the O$^{++}$ that there would be significant
   [Ne~V] and [Fe~VII] emission; specifically, U = 10$^{-1.3}$, which, given the
   derived luminosity in ionizing photons, sets the distance from the
   central source at 15 parsecs. The four pointings in this set of data span
   a region of $\sim$ 70 parsec from the central source. Given 
   the uncertainty in the actual location of the source and the value of the
   fraction of reflected continuum (Pier et al. 1994), the choice of distance
   is plausible and puts our model region right in the middle of the set
   of FOS pointings. At this distance, the high ionization component is 
   characterized by an ionization parameter, U = 10$^{-1}$. 

   It is clear from the SED of the ionizing continuum and the observed
   He~II $\lambda$4686/H$\beta$ ratio that much of the gas in these regions is optically
   thin at the Lyman limit. For optically thick gas, a simple photon counting
   calculation (cf Kraemer et al. 1994) yields a 4686/H$\beta$ ratio $\sim$ 0.16; 
   observed values range from 0.43 to 0.59. The presence of a large fraction of 
   optically thin (matter bounded) gas would increase the relative strength of 
   the highest ionization lines in the composite spectrum, compared to the 
   composite spectrum from a region composed entirely of optically thick 
   (radiation bounded) gas. This is supported by the fact that the strongest 
   relative [Ne~V] $\lambda$3426 emission is seen in the same region that has the 
   strongest He~II $\lambda$4686. There is no definitive way to determine the exact 
   optical (and physical) thickness of each component of emission line gas in 
   these regions. We decided, a priori, to truncate the integration in the O$^{
++}$ 
   component at an optical depth of 10 at the Lyman limit. The resulting 
   filtered spectrum is shown in Figure 5, and shows complete absorption at the He~II 
   Lyman limit, and strong absorption at the hydrogen Lyman limit; the physical 
   conditions in gas photoionized by this continuum will be strongly effected 
   by X-ray ionization and heating processes, as noted above. The resulting
   ionization parameter for the shielded gas was 10$^{-2.35}$, with most of the 
   energy in X-rays. We truncated the integration for the shielded 
   component when the ionized fraction of the gas dropped below 5\% and there 
   was no longer significant line emission generated other than [N~I] $\lambda$5200 and 
   [O~I] $\lambda\lambda$6300,6364, neither of which are strong in these data.
   
   Since the relative contributions from the O$^{++}$ and shielded components
   are linked, there is at least some basis for truncating the integration
   of the O$^{++}$ at the chosen optical depth. Picking the point to truncate the
   integration of the high ionization component is somewhat more arbitrary.
   We chose to truncate the model when we reached an optical depth of 10 at the 
   He~II Lyman limit, although there are indications from the predicted line
   ratios this component might be even thinner.

   The results of the three component models are given in Table 3, along with
   the composite spectrum line ratios and a dereddened "observed" spectrum,
   averaged over the four sets of observations. The relative contributions to 
   the composite spectrum are as follows: 25\% from the high ionization component, 50\% from 
   the O$^{++}$ component and 25\% from the shielded component. We can check
   the plausibility of these ratios by comparing the total H$\beta$ emission from
   each component. We only know one dimension of these slabs: the physical depth
   from the ionized face to the point where we truncated the integration.
   Comparing the product of the physical depth and the average emissivity gives
   a measure of the possible contribution from each component. The ratio of this
   product for the O$^{++}$ and high ionization component is $\sim$2, which implies 
   comparable covering factors for each of these components and supports the
   ratio of relative contribution used for the composite spectrum. The ratio of
   the products from the O$^{++}$ and shielded components is $\sim$4, or approximately
   twice the ratio used in the composite. Since these two components are
   restricted to the same covering factor, as described above, it would
   seem that we do not have enough of the shielded gas. One obvious explanation
   is that there is additional shielded gas 'behind' the high ionization 
   component. There may be other explanations as well; we will address this 
   in the following section.
   
   Comparison of the composite line ratios with the averaged observed ratios
   in Table 3
   shows good agreement for the majority of emission lines. In particular, 
   these include the [Fe~VII] lines, [Ne~V] $\lambda$3426, 
   [Ne~III] $\lambda$3869, [N~II] $\lambda\lambda$6584,6584 [O~II] $\lambda$3727, 
   [S~II] $\lambda\lambda$6716,6731, C~IV $\lambda$1550, and N~V $\lambda$1240. The
   fact that the model predictions are good for lines of such a wide 
   range of ionization state indicates that our overall balance of
   high and low excitation gas is reasonable. Also, the model predictions
   support our assumptions about the elemental abudances, since relative
   strengths of the iron, neon and nitrogen lines would all decrease if solar
   abundances were assumed for these elements. The [O~III] $\lambda$5007/$\lambda$4363 ratio
   is also in good agreement with the observations, but this was to be 
   expected since the choice of density and ionization parameter for the
   O$^{++}$ component was based on this ratio. The predicted ratio of the 
   [O~I] $\lambda$6300/([O~I] $\lambda$6364~+~[Fe~X] $\lambda$6374) is also in agreement with the observations, 
   showing that the 6364 line is indeed blended with [Fe~X], although the
   strength of these lines relative to H$\beta$ is somewhat high. The 
   predictions of the ratios of the He~II lines to H$\beta$ are also in good 
   agreement with the observations, which support our inclusion of matter 
   bounded gas in the composite model. 
   
   From the model predictions we can determine the size of the emitting
   regions and the total amount of excited gas, using the relative 
   contributions given above and comparing them to total H$\beta$ emission from 
   the nucleus.  Assuming the distance to NGC1068 is 20 Mpc and a filling
   factor within the emitting region of unity we obtain a minimum volume 
   of $\sim$1 pc$^{3}$; this is quite reasonable given the 20 parsec extent
   of each region their apparent clumpiness (see Figure 1). We
   compute an actual filling factor of $\sim$ 10$^{-4}$. The total mass of gas 
   required is $\sim$ 8,000$M\odot$, which is not unreasonable within such a volume.

   Although we were able to obtain a reasonable fit to the data with this
   model, we would not suggest that this is a unique but, rather,
   possible solution. In addition to the problem of emissivity and
   covering factor of the shielded component, there are several discrepant
   emission line ratios, which we will address in the following section.
   However, from the success of this model, we can, with some confidence,
   make several statements about the physical conditions in these regions.
   First, the dominant source of ionization and heating is photoionization
   from the continuum radiation emitted by the central source. The estimate of 
   the intrinsic SED and luminosity of the central source by Pier et al. (1994)
   is approximately correct, including the reflection fraction. There is
   a range of density within the emission line gas, and some of the
   gas is dusty. And, finally, it is likely that the elemental abundances
   in these regions are not solar.

\section{Model Discrepancies}

   As discussed in section 2., the data obtained with FOS are spatially
   resolved. This permitted us to adjust more parameters, in particular
   abundances, than we have in previous studies (cf Kraemer et al. 1998). 
   The result of this flexibility is a better set of predicted line
   ratios than can usually be achieved with photoionization models, but there
   are still obvious discrepancies. These both show the limitations
   of these simple models and can be used to obtain additional physical 
   insight. Although there are some weak lines which were not well fit
   by the models (e.g. [Mg~V] $\lambda$2929 and [Ne~V] $\lambda$2974), the
   fluxes of the weakest lines were difficult to measure accurately. We
   will, therefore, concentrate on the discrepancies in the predicted
   strengths of the stronger lines. 

   There are three high ionization UV lines for which the model predictions are most
   obviously discrepant: N~IV] $\lambda$1485, O~III] $\lambda$1664 and [Ne~IV] $\lambda$2424. The former
   two are predicted too strong, by factors of 4 and 3 respectively. The
   [Ne~IV] line is predicted too weak by a factor of 4. The N~IV] line is a strong
   coolant in both of the directly ionized components of the model. If the
   high ionization component was truncated at lower optical depth or was
   characterized by a higher ionization parameter, the N$^{+3}$ zone would be 
   smaller, reducing the strength of the $\lambda$1485 line. This would also help
   reduce the relative strength of the O~III] $\lambda$1664 line. The problem is that
   it would worsen the fit for the [Ne~V] and [Fe~VII] lines. The weakness of
   the predicted [Ne~IV] strength presents a different problem. With the
   SED assumed for these models, it takes a unique set of conditions to 
   get a component in which the relative [Ne~IV] strength is comparable to
   that observed. The contribution from such a component would be diluted
   by the other component spectra, still resulting in a relatively weak 
   $\lambda$2424 line.
   One possible explanation is that the electron temperatures predicted
   by these models is too low. A higher electron temperature in the O$^{++}$
   component would increase the strengths of the collisionally excited lines.
   Combined with the more highly ionized, high ionization component, the 
   overall strength of the [Fe~VII] and [Ne~V] lines could be maintained,
   and [Ne~IV] increased, while dropping the overall the N~IV] and O~III].
   Although at first glance this might present a problem for the [O~III] $\lambda$5007/
   $\lambda$4363 ratio, the increase in the relative 4363 strength could be offset
   by lowering the density of the O$^{++}$ component. The question is: what is
   the source of this additional heating? We shall return to this question
   in the following section.
   
   There are also discrepancies in the low excitition lines, which arise 
   primarily in the shielded component. The most obvious problem is that the 
   reddening corrected Balmer decrement is much steeper than predicted by the 
   models, although this is biased somewhat in the average of the observations 
   by the extremely high N$\alpha$/H$\beta$ ratio seen in the Position 3 
   spectrum. The predicted neutral oxygen lines are too strong, 
   and the model predicts fairly strong [N~I] emission, which we did not
   detect in the spectra. The Mg~II $\lambda$2800 line is too weak
   by approximately a factor of 3. Finally, the either the emissivity or size 
   of the partially ionized zone in the shielded component is insufficient. 
   If more ionizing energy were injected into this gas, all of these 
   discrepancies would be mitigated. A larger ionization fraction would
   drop the relatively strength of the neutral lines. Increased heating and
   ionization would increase the emissivity of the gas. And, if the
   energy injection were in the form of energetic particles or increased
   X-ray ionization, enhanced collisional exciation of neutral hydrogen
   would increase the ratio of H$\alpha$/H$\beta$.

\section{Discussion}

   Although photoionization by the central source is the
   dominant mechanism determining the physical conditions in these four
   regions, these simple models do not give the full picture of the
   underlying physics in the emission line gas. There appears to be
   additional heating and/or ionization from some source other than the
   assumed ionizing continuum. 

   There is evidence that emission line ratios in these regions are
   affected by processes other than pure photoionization. Kriss et al. (1992)
   have remarked on the surprising strength of the N~III $\lambda$990 and C~III $\lambda$977 
   resonance lines seen in HUT spectra of NGC 1068. Although these lines are 
   outside the FOS bandpass, our model predictions of their strengths 
   relative to the intercombination lines, N~III] $\lambda$1785 and C~III] $\lambda$1909, are well 
   below those measured in the HUT data (although we note that the HUT data
   area smapling a much larger region). Kriss et al. found similar problems with
   the model predictions for these lines and attributed their
   relative enhancement as due to additional heating of the emission line 
   gas by shocks. Ferguson et al. (1995) have suggested that the enhancement is 
   not due to shock heating but, rather, continuum fluorescence. Once 
   a driver line is excited by absorption of a continuum photon, it will scatter and
   degrade into subordinate lines, resulting in enhancement of the resonance
   lines similar to that seen for the Lyman transitions in the Case B
   approximation for hydrogen (Osterbrock 1989). This process is only 
   important for optically thick driver lines and is dependent on the line width and, 
   therefore, the turbulent velocity in the emission line gas. Although
   it is likely that the conditions within the line formation regions are
   favorable for this process, it is difficult to determine how much
   enhancement there will be. For example, Ferguson et al. get only about half
   the observed value for Doppler widths $\sim$ 1000 km/sec.  Furthermore, the 
   strengths of lines unaffected by continuum fluorescence are also 
   underpredicted by the models. Therefore, it is likely that some type
   of additional heating is required.

   Lame \& Ferland (1991) have examined the relative contributions of 
   photoionization and shocks to the excitation of the emission line gas in 
   the planetary nebula NGC 6302. Their models of gas photoionization by a 
   central star of temperature 450,000K accurately reproduced the relative 
   strengths of most nebular lines, they could not match the observed
   strengths of the highest ionization lines, such 
   as N~IV] $\lambda$1485, N~V $\lambda$1240, C~IV $\lambda$1549, 
   [Ne~IV] $\lambda$ 2424, and [Ne~V] $\lambda$3426. Their 
   conclusion was that the physical conditions in the nebula were likely to be 
   the result of photoionization combined with collisional excitation from 
   shocks due to stellar winds. It seems, at first, that such a combination of 
   effects could resolve some of the discrepancies seen with our simple 
   photoionzation models.
   
   If the additional heating is due to shocks, Kriss et al. (1992) suggest as
   one possible mechanism the interaction of the radio plasma ejected from the
   nucleus with the NLR clouds. This was first proposed by Wilson \&
   Ulvestad (1983) after VLA maps of the radio emission in NGC 1068 revealed
   the existence of a jet within the NLR. The coincidence of the brightest [O~III] knots
   with the radio bright spots (Evans et al. 1991) and recent FOC spectra
   showing possible jet driven motions in the NLR (Axon et al. 1998) support
   this scenario. As shown in Figure 1, the four regions discussed here
   all lie on or near the axis of the radio source, with Position 3 lying
   directly in its projected path. Although Position 3 is probably
   furthest from the ionizing source, its spectrum not 
   only has the strongest high excitation lines (which may only be indicative of a 
   greater fraction of optically thin gas), but the smallest 
   [O~III] $\lambda$5007/$\lambda$4363, indicating the highest electron temperature in the O$^{++}$
   zone (assuming density similar to the other regions).

   It then seems plausible that the additional heating we infer from our models
   may be the result of the result of a cloud/jet interaction. Is it less
   certain that the additional energy is the result of shock heating. 
   Although it is likely that an additional heating source is present
   in the most highly ionized gas in these regions, it is also apparent
   from the model discrepancies that additional heating and ionization are
   required in the shielded gas. Ferland \& Mushotzky (1984) have discussed
   the effects of the injection of relativistic electrons into emission-line
   gas for conditions applicable to the NLR of AGN. Their models predict
   an increase in the temperature, ionization fraction, and physical size of 
   partially ionized or neutral gas deep within a photoionized emission line 
   cloud, as well as the temperature and ionization fraction of the illuminated
   face of the cloud. In their simple model, relativistic electrons
   were able to penetrate up to column densities in excess of 10$^{22}$cm$^{-2}$
   , similar to the sizes of our model components. The plausibility of this 
   explanation is supported by the fact that there is clearly a source for 
   such energetic  particles in the inner NLR, i.e. the radio jet, and that 
   the spectrum of Position 3 shows the largest H$\alpha$/H$\beta$ ratio.

\section{Conclusions}

   We have analyzed UV and optical spectra of the Seyfert 2 galaxy NGC 1068,
   obtained with the FOS on {\it HST}, from four regions within the inner NLR. 
   Although these data were taken before the installation of COSTAR, we were 
   able to determine that the contamination from the aberrated PSF did not 
   significantly degrade the quality of the data obtained from two of the four 
   pointings, providing us with three relatively distinct regions. We have 
   constructed photoionization models to match an averaged set of conditions 
   from these regions. The predicted emission line ratios fit the dereddened 
   observed ratios for the large majority of emission lines, with the few 
   exceptions noted in section 6. We were able to fit both permitted and 
   forbidden lines and lines from a wide range of ionization state, e.g. N~V 
   as well as [N~II], using a three-component model and a limited set of free 
   parameters. Since these models are constrained by the best estimate of the 
   underlying SED of the ionizing continuum and spatial information provided 
   by the FOS data, we are confident that the general physical characteristics
   assmed in these models reflect the actual physical conditions in the
   NLR gas.

   From our analysis and modeling of the spectra we can make several
   statements regarding the physical conditions in the inner NLR of
   NGC 1068. First of all, the dominant mechanism for ionizing the 
   NLR gas is photoionization by continuum radiation from the central
   source. The estimates by Pier et al. (1994) and Miller et al. (1991) 
   regarding the SED and intrinsic luminosity of the ionizing continuum
   are approximately correct. As noted by Netzer (1997) and Netzer \&
   Turner (1997), the abundances in the emission line gas are not
   solar, although we found that enhancement of nitrogen was more likely
   than depletion of oxygen. Also, our models support Netzer's suggestion 
   that these regions include a mixture of dusty and dust-free gas. 
   
   There are two apects of these models that provide us with additional
   insight into the physical conditions in the NLR. First, as we discussed
   in detail in the previous two sections, the electron temperatures
   and, perhaps, ionization states predicted by the models are too low,
   although this is masked somewhat by our choice of initial conditions
   such as density. The most likely explanation, given the constraints
   on the SED, is that there is additional collisional heating and ionization.
   Kriss et al. (1992) have suggested that shock heating in an important process.
   Since additional heating and ionization are also needed in the partially
   ionized component, it may be that cosmic rays, perhaps associated
   with the radio jet, are the main source of additional energy. As shown
   by Ferland \& Mushotzky (1984), relativistic electrons can penetrate
   through large column densities of atomic gas, affecting the extended
   partially ionized evelope. The location of the four regions in our data
   set with respect to the radio jet, particularly Position 3, supports
   the suggestion of a jet/cloud interaction, whether the additional heating
   is due to shocks, comsic rays, or some combination of the two.

   The other interesting aspect of the model is that the most likely source of the
   low ionization emission lines is gas that is partially screened
   by an optically thin component, probably of higher density. In fact,
   if there is additional heating beyond that due to photoionization,
   the ratio of the densities of the screening component to the low ionization
   gas may be even greater than assumed here (2:1). We would suggest
   that this density gradient is the result of material being swept up
   by the force of the jet, creating a thin wall of denser material nearest
   the ionizing source. This is in general agreement with the recent 
   observations by Axon et al. (1998), which indicate that the gas with the 
   strongest line emission lies along the direction of the radio jet, although
   this may be due, in part, to the collimation of the ionizing radiation.

   Planned GTO and GO observations of NGC 1068 with the Space Telescope 
   Imaging Spectrograph (STIS), on {\it HST}, will provide optical and UV spectra
   data with improved spatial resolution over a larger section of the NLR. 
   Better spatial resolution will permit us to examine conditions within
   these regions and may reveal more about their apparent inhomogeneity.
   From these data we can further examine the possibility of interaction 
   between the radio jet and the ionized gas, including looking off the radio
   axis for NLR gas that may be unaffected by the jet. Such obervations
   will go further in constraining the physical conditions and energy
   budget in the NLR.

\acknowledgments

\clearpage

\figcaption[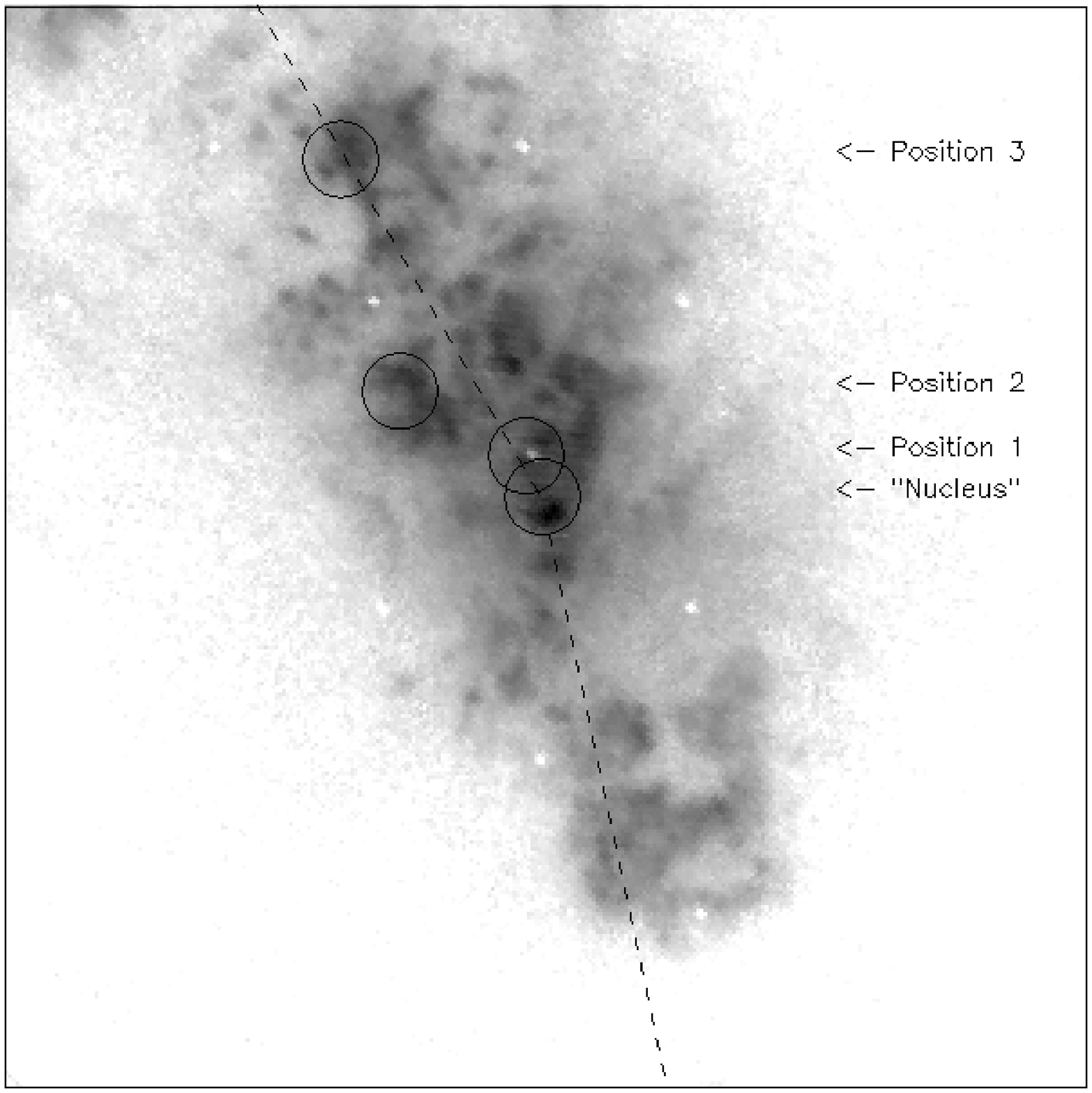]{FOC [O III] image of NGC 1068 with the four FOS aperture 
positions. The square pattern of spots with low counts are instrumental reseau 
marks. North is up and east is to the left. The apertures are 0\arcsecpoint3 in 
diameter. 
}\label{fig1}

\figcaption[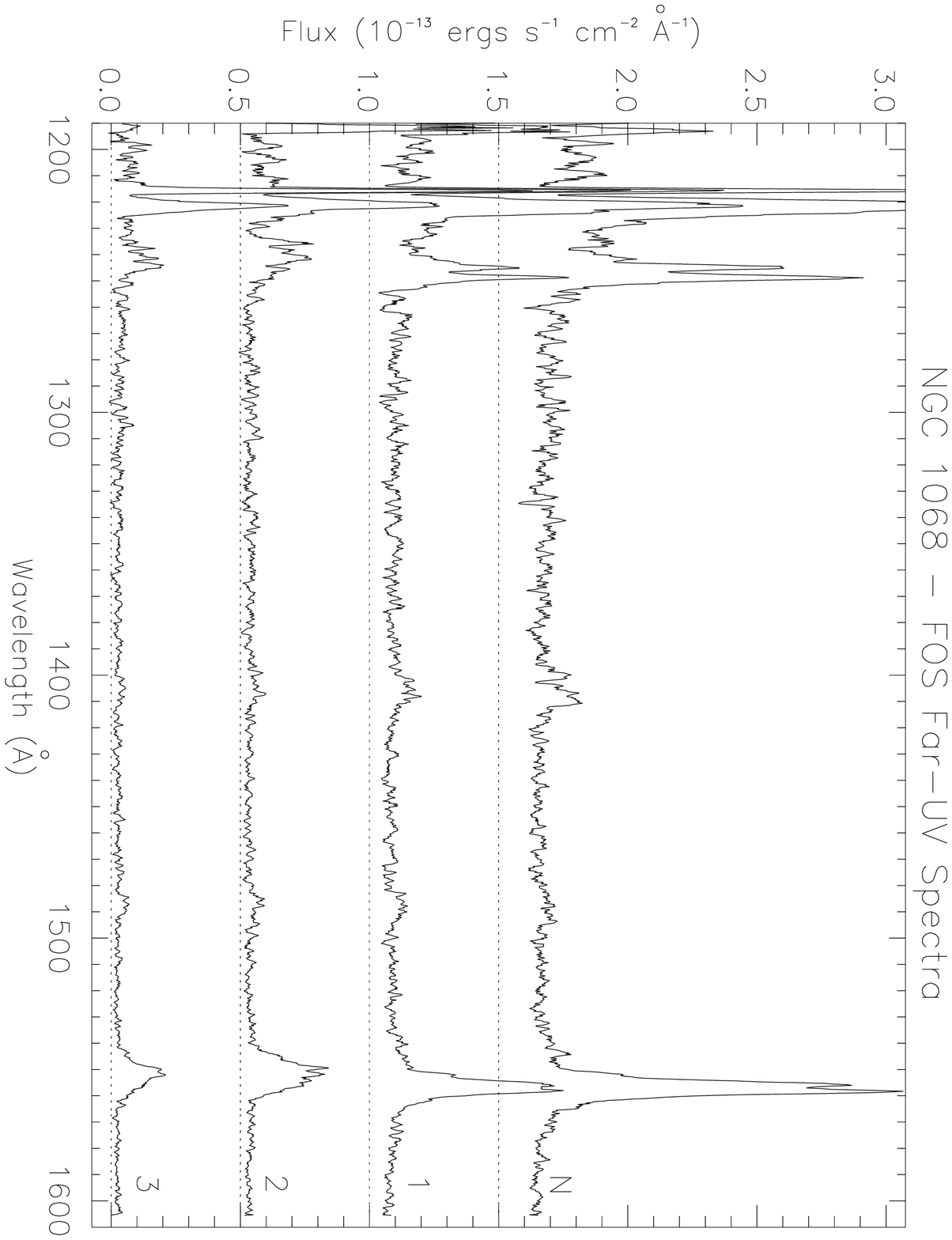]{FOS Far-UV (G130H) spectra of NGC 1068 for the nucleus 
(``N'') and positions 1, 2, and 3).
}\label{fig2} 

\figcaption[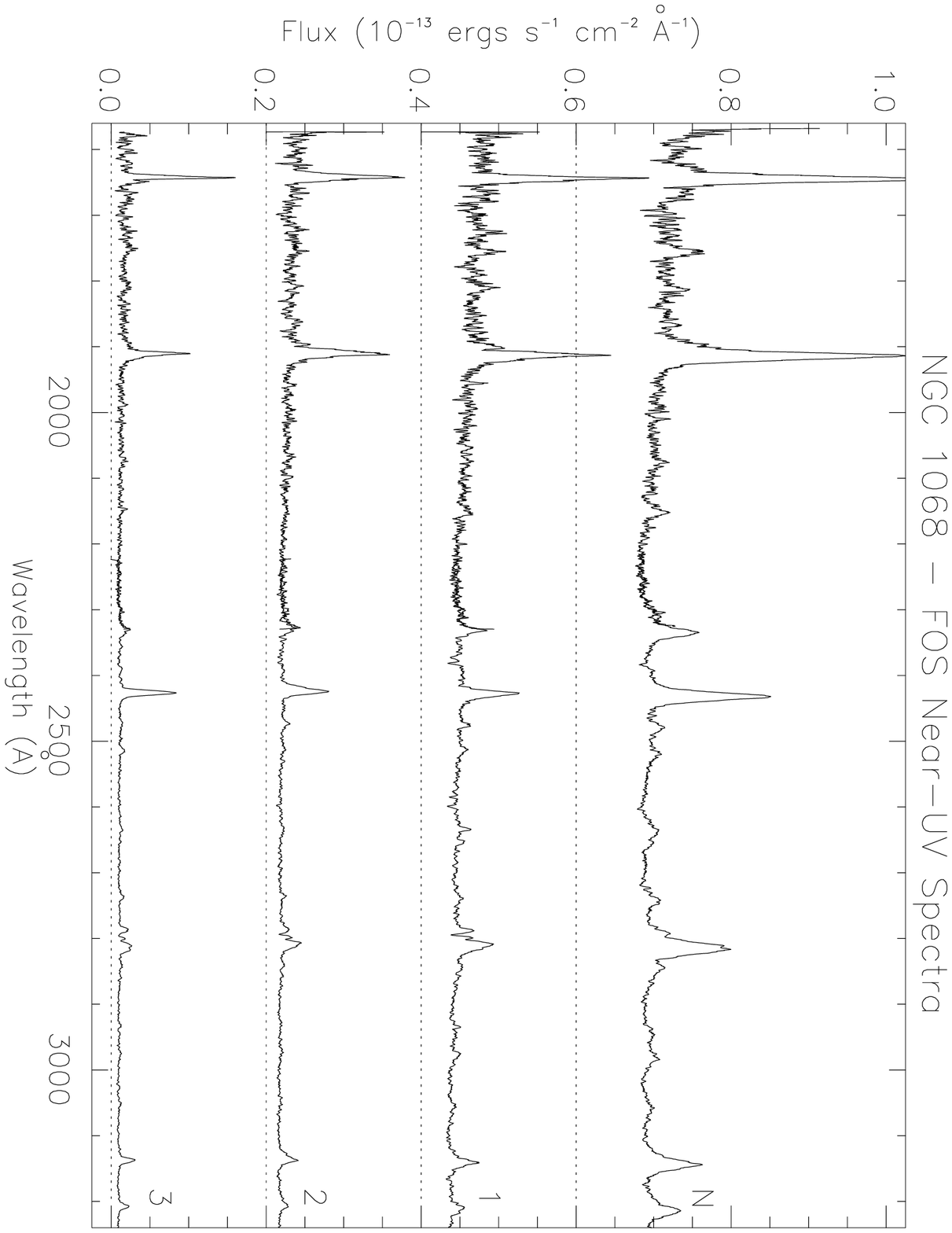]{FOS Near-UV (G190H, G270H) spectra of NGC 1068 for the 
nucleus (``N'') and positions 1, 2, and 3).
}\label{fig3} 

\figcaption[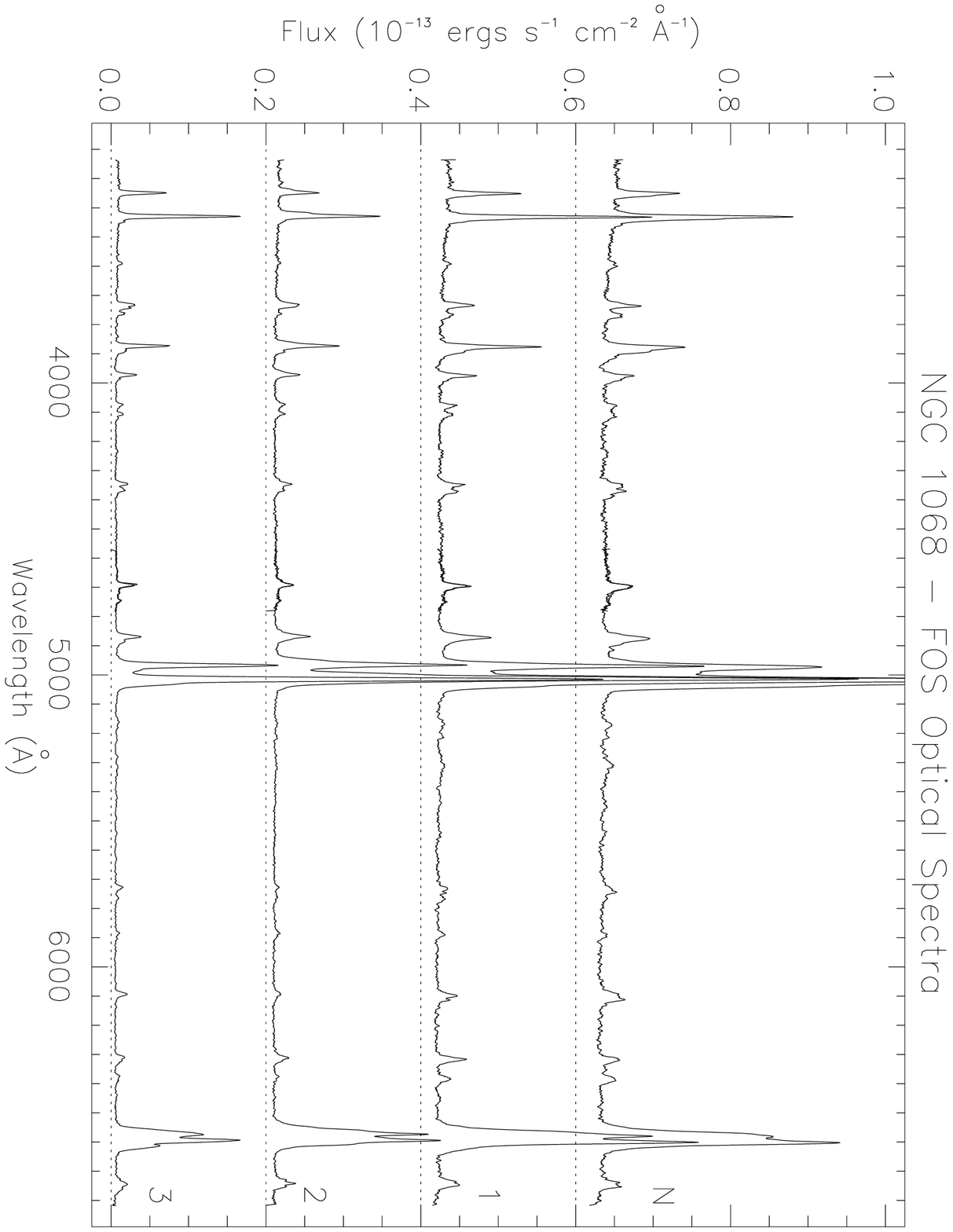]{FOS Optical (G400H, G570H) spectra of NGC 1068 for the 
nucleus (``N'') and positions 1, 2, and 3).
}\label{fig4} 

\figcaption[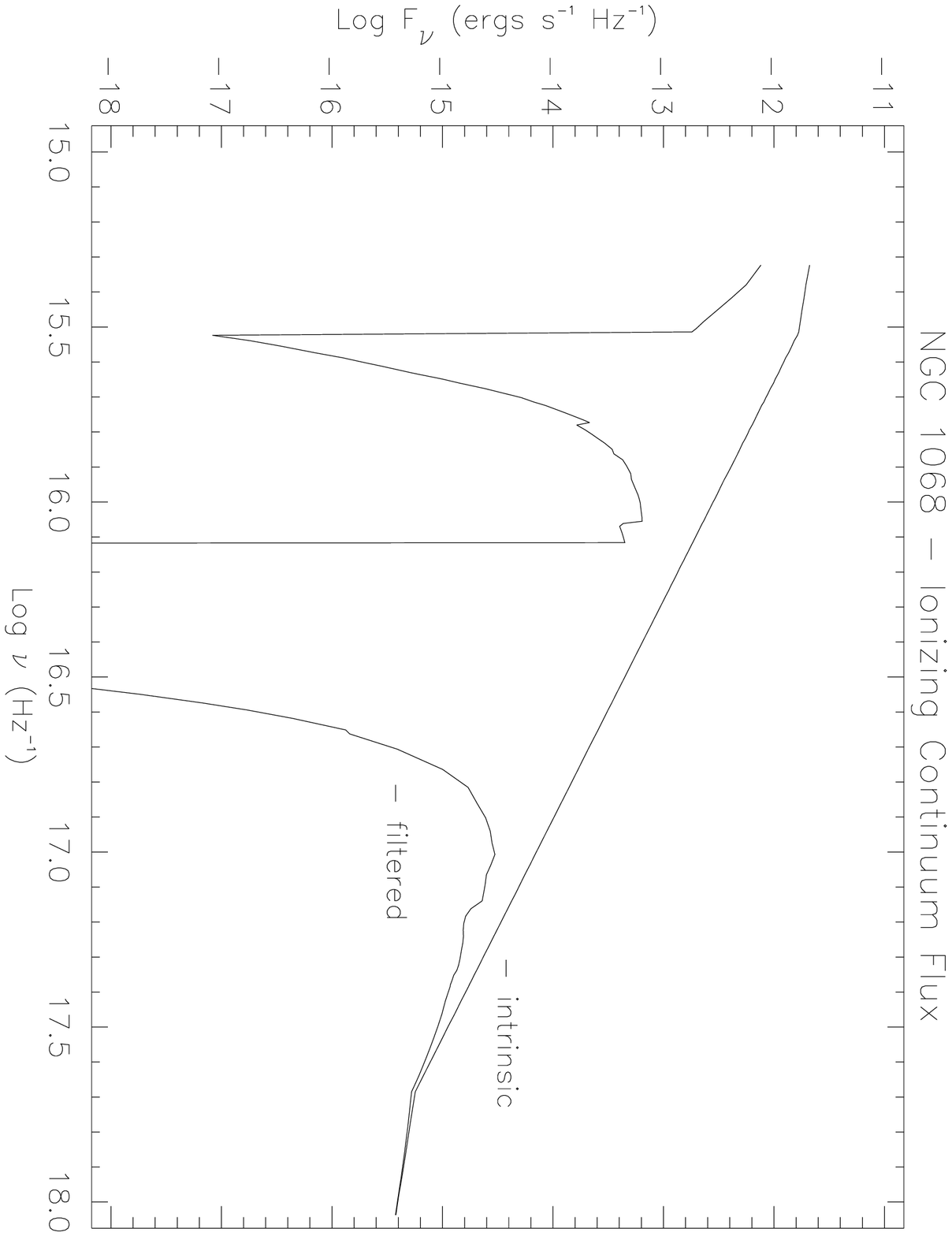]{Comparison of incident ionizing flux spectrum at the 
illuminated face of a directly photoionized cloud to the filtered flux spectrum used for the shielded 
model component. 
}\label{fig5}

\clearpage
\clearpage
\begin{deluxetable}{lccccc}
\tablecolumns{6}
\footnotesize
\tablecaption{Log of FOS observations (0\arcsecpoint3 diameter aperture)
\label{tbl-1}}
\tablewidth{0pt}
\tablehead{
\colhead{Archive} & \colhead{Dataset} & \colhead{Detector} & 
\colhead{Grating} & \colhead{Exposure} & \colhead{Observation} \\
\colhead{Name} & \colhead{Name} & \colhead{} & 
\colhead{} & \colhead{(sec)} & \colhead{Date}
}
\startdata
NGC1068-NUC	& Y0GQ0107T  & FOS/BL  & G130H  & 1500 & 1991 January 27 \\
NGC1068-NUC	& Y0GQ0106T  & FOS/BL  & G190H  & 1000 & 1991 January 27 \\
NGC1068-NUC	& Y0GQ0108T  & FOS/BL  & G270H  &  700 & 1991 January 27 \\
                &	     &         &	&	    & \\  
NGC1068 	& Y0MW0707T  & FOS/BL  & G130H  & 1500 & 1991 June 25 \\
NGC1068 	& Y0MW0706T  & FOS/BL  & G190H  & 1000 & 1991 June 25 \\
NGC1068 	& Y0MW0806T  & FOS/RD  & G270H  &  700 & 1991 June 25 \\
NGC1068 	& Y0MW0809T  & FOS/RD  & G400H  &  600 & 1991 June 25 \\
NGC1068 	& Y0MW0807T  & FOS/RD  & G570H  &  600 & 1991 June 25 \\
                &	     &         &	&	   &	  \\
NGC1068-CLOUD1  & Y0MW070AT  & FOS/BL  & G130H  & 1500 & 1991 June 25 \\
NGC1068-CLOUD1  & Y0MW0709T  & FOS/BL  & G190H  & 1000 & 1991 June 25 \\
NGC1068-CLOUD1  & Y0MW080BT  & FOS/RD  & G270H  &  700 & 1991 June 25 \\ 
NGC1068-CLOUD1  & Y0MW080ET  & FOS/RD  & G400H  &  600 & 1991 June 26 \\
NGC1068-CLOUD1  & Y0MW080CT  & FOS/RD  & G570H  &  600 & 1991 June 26 \\
                &	     &         &	&	   &	  \\
NGC1068-CLOUD2  & Y0MW070DT  & FOS/BL  & G130H  & 1500 & 1991 June 25 \\
NGC1068-CLOUD2  & Y0MW070CT  & FOS/BL  & G190H  & 1000 & 1991 June 25 \\
NGC1068-CLOUD2  & Y0MW080GT  & FOS/RD  & G270H  &  700 & 1991 June 26 \\
NGC1068-CLOUD2  & Y0MW080JT  & FOS/RD  & G400H  &  600 & 1991 June 26 \\
NGC1068-CLOUD2  & Y0MW080HT  & FOS/RD  & G570H  &  600 & 1991 June 26 \\
                &	     &         &	&	   &	  \\
NGC1068-CLOUD3  & Y19G0206T  & FOS/BL  & G130H  & 2000 & 1993 March 4 \\
NGC1068-CLOUD3  & Y19G0205T  & FOS/BL  & G190H  & 1350 & 1993 March 3 \\
NGC1068-CLOUD3  & Y19G0107T  & FOS/RD  & G270H  & 1000 & 1993 March 1 \\
NGC1068-CLOUD3  & Y19G0106T  & FOS/RD  & G400H  &  850 & 1993 March 1 \\
NGC1068-CLOUD3  & Y19G0105T  & FOS/RD  & G570H  &  850 & 1993 March 1 \\
\enddata
\end{deluxetable}

\clearpage
\begin{deluxetable}{lrrrr}
\tablecolumns{5}
\footnotesize
\tablecaption{Dereddened Line ratios from FOS positions 
(relative to H$\beta$)\label{tbl-2}}
\tablewidth{0pt}
\tablehead{
\colhead{} & \colhead{Nucleus} & \colhead{Position 1} & \colhead{Position 2} & 
\colhead{Position 3} 
}
\startdata
Ly$\alpha$ $\lambda$1216                     &11.83 ($\pm$1.11) &18.91 
($\pm$2.68)   &12.03 ($\pm$2.88)   &16.66 ($\pm$3.95)  \\
N V $\lambda$1240           	      	     & 4.86 ($\pm$0.38) &10.74 
($\pm$1.72)   & 7.87 ($\pm$1.65)   & 8.78 ($\pm$2.52)  \\
Si IV $\lambda$1398/ O IV] $\lambda$1402     & 1.30 ($\pm$0.22) & 2.09 
($\pm$0.47)   & 2.38 ($\pm$0.57)   & 1.55 ($\pm$0.62)  \\
N IV] $\lambda$1486         	      	     & 0.47 ($\pm$0.15) & 1.39 
($\pm$0.31)   & 1.80 ($\pm$0.43)   & 1.65 ($\pm$0.31)  \\
C IV $\lambda$1550          	      	     & 6.93 ($\pm$0.23) & 9.72 
($\pm$0.96)   & 8.65 ($\pm$1.07)   & 9.60 ($\pm$1.60)  \\
He II $\lambda$1640         	      	     & 3.20 ($\pm$0.24) & 3.05 
($\pm$0.35)   & 3.06 ($\pm$0.43)   & 4.26 ($\pm$0.73)  \\
O III] $\lambda$1663        	      	     & 0.36 ($\pm$0.10) & 0.28 
($\pm$0.16)   & 0.22 ($\pm$0.13)   & 0.61 ($\pm$0.11)  \\
$\lambda$1718?		        	     & -----~~~~~~~~~~~ & 0.44 
($\pm$0.16)   & 0.36 ($\pm$0.18)   & 1.17 ($\pm$0.19)  \\
N III] $\lambda$1750                  	     & 0.38 ($\pm$0.05) & 0.48 
($\pm$0.15)   & 0.67 ($\pm$0.34)   & 0.81 ($\pm$0.23)  \\
$\lambda$1804?				     & 0.21 ($\pm$0.05) & 0.55 
($\pm$0.18)   & 0.67 ($\pm$0.36)   & 0.80 ($\pm$0.29)  \\
C III] $\lambda$1909, Si III] $\lambda$1892  & 3.99 ($\pm$0.25) & 3.80 
($\pm$0.41)   & 4.78 ($\pm$0.52)   & 5.04 ($\pm$1.01)  \\
$\lambda$2143?				     & 0.23 ($\pm$0.02) & 0.44 
($\pm$0.12)   & 0.72 ($\pm$0.21)   & 0.68 ($\pm$0.30)  \\
$[$O III] $\lambda$2321/C II] $\lambda$2326  & 1.51 ($\pm$0.15) & 1.18 
($\pm$0.18)   & 0.91 ($\pm$0.14)   & 0.88 ($\pm$0.24)  \\
$[$Ne IV] $\lambda$2423     	      	     & 1.97 ($\pm$0.21) & 1.72 
($\pm$0.17)   & 1.66 ($\pm$0.18)   & 3.25 ($\pm$0.54)  \\
$[$O II] $\lambda$2470      	      	     & 0.27 ($\pm$0.07) & 0.20 
($\pm$0.04)   & 0.17 ($\pm$0.04)   & 0.37 ($\pm$0.14)  \\
He II $\lambda$2512			     & 0.27 ($\pm$0.05) & 0.22 
($\pm$0.04)   & 0.14 ($\pm$0.04)   & 0.22 ($\pm$0.05)  \\
$\lambda$2628?				     & 0.49 ($\pm$0.05) & 0.55 
($\pm$0.06)   & 0.23 ($\pm$0.08)   & 0.31 ($\pm$0.08)  \\
He II $\lambda$2734			     & 0.37 ($\pm$0.03) & 0.28 
($\pm$0.06)   & 0.16 ($\pm$0.06)   & 0.24 ($\pm$0.03)  \\
Mg II $\lambda$2800         	      	     & 1.97 ($\pm$0.10) & 1.28 
($\pm$0.18)   & 1.01 ($\pm$0.10)   & 1.11 ($\pm$0.11)  \\
$[$Mg V] $\lambda$2929         	      	     & 0.21 ($\pm$0.02) & 0.22 
($\pm$0.04)   & 0.11 ($\pm$0.02)   & 0.08 ($\pm$0.02)  \\
$[$Ne V] $\lambda$2974                	     & 0.28 ($\pm$0.05) & 0.25 
($\pm$0.03)   & 0.14 ($\pm$0.03)   & 0.12 ($\pm$0.03)  \\
$\lambda$3045?			             & 0.29 ($\pm$0.06) & 0.27 
($\pm$0.07)   & 0.15 ($\pm$0.06)   & 0.17 ($\pm$0.08)  \\
O III $\lambda$3133			     & 0.95 ($\pm$0.02) & 0.66 
($\pm$0.04)   & 0.56 ($\pm$0.04)   & 0.63 ($\pm$0.12)  \\
He II $\lambda$3204                   	     & 0.27 ($\pm$0.04) & 0.29 
($\pm$0.05)   & 0.25 ($\pm$0.04)   & 0.36 ($\pm$0.04)  \\
$[$Ne V] $\lambda$3346      	      	     & 0.81 ($\pm$0.02) & 1.29 
($\pm$0.05)   & 1.02 ($\pm$0.04)   & 1.66 ($\pm$0.10)  \\
$[$Ne V] $\lambda$3426      	      	     & 2.32 ($\pm$0.05) & 3.45 
($\pm$0.11)   & 2.67 ($\pm$0.09)   & 4.29 ($\pm$0.22)  \\
\tablebreak
$[$Fe VII] $\lambda$3588    	      	     & 0.19 ($\pm$0.03) & 0.15 
($\pm$0.02)   & 0.17 ($\pm$0.03)   & 0.20 ($\pm$0.03)  \\
$[$O II] $\lambda$3727      	      	     & 0.46 ($\pm$0.02) & 0.55 
($\pm$0.05)   & 0.83 ($\pm$0.04)   & 1.25 ($\pm$0.06)  \\
$[$Fe VII] $\lambda$3760    	      	     & 0.37 ($\pm$0.02) & 0.26 
($\pm$0.04)   & 0.18 ($\pm$0.03)   & 0.32 ($\pm$0.02)  \\
$[$Ne III] $\lambda$3869    	      	     & 2.46 ($\pm$0.13) & 2.41 
($\pm$0.08)   & 2.02 ($\pm$0.10)   & 2.51 ($\pm$0.09)  \\
$[$Ne III] $\lambda$3967              	     & 0.72 ($\pm$0.04) & 0.85 
($\pm$0.02)   & 0.81 ($\pm$0.06)   & 0.91 ($\pm$0.06)  \\
$[$S II] $\lambda$4072                	     & 0.28 ($\pm$0.03) & 0.31 
($\pm$0.01)   & 0.35 ($\pm$0.03)   & 0.30 ($\pm$0.18)  \\
H$\delta$ $\lambda$4100 	      	     & 0.30 ($\pm$0.03) & 0.32 
($\pm$0.02)   & 0.30 ($\pm$0.05)   & 0.32 ($\pm$0.02)  \\
H$\gamma$ $\lambda$4340      	      	     & 0.47 ($\pm$0.01) & 0.56 
($\pm$0.03)   & 0.53 ($\pm$0.03)   & 0.53 ($\pm$0.06)  \\
$[$O III] $\lambda$4363      	      	     & 0.51 ($\pm$0.04) & 0.45 
($\pm$0.03)   & 0.34 ($\pm$0.07)   & 0.54 ($\pm$0.08)  \\
He II $\lambda$4686          	      	     & 0.44 ($\pm$0.03) & 0.42 
($\pm$0.01)   & 0.42 ($\pm$0.03)   & 0.59 ($\pm$0.06)  \\
H$\beta$                 	      	     & 1.00 	           & 1.00 	 
    & 1.00          	 & 1.00    	      \\
$[$O III] $\lambda$5007                      &13.14 ($\pm$0.07) &14.85 
($\pm$0.19)   &15.40 ($\pm$0.21)   &17.11 ($\pm$0.14)  \\
$[$Fe VII] $\lambda$5721     	      	     & 0.48 ($\pm$0.11) & 0.37 
($\pm$0.04)   & 0.25 ($\pm$0.03)   & 0.38 ($\pm$0.03)  \\
He I $\lambda$5876           	      	     & 0.13 ($\pm$0.01) & 0.14 
($\pm$0.02)   & 0.27 ($\pm$0.04)   & 0.17 ($\pm$0.04)  \\
$[$Fe VII] $\lambda$6087     	      	     & 0.71 ($\pm$0.09) & 0.48 
($\pm$0.03)   & 0.27 ($\pm$0.02)   & 0.47 ($\pm$0.05)  \\
$[$O I] $\lambda$6300        	      	     & 0.41 ($\pm$0.04) & 0.50 
($\pm$0.02)   & 0.42 ($\pm$0.03)   & 0.47 ($\pm$0.03)  \\
$[$O I] $\lambda$6364/$[$Fe X] $\lambda$6374 & 0.37 ($\pm$0.15) & 0.31 
($\pm$0.03)   & 0.12 ($\pm$0.02)   & 0.20 ($\pm$0.02)  \\
$[$N II] $\lambda$6548, 6584         	     & 4.45 ($\pm$0.62) & 4.86 
($\pm$0.34)   & 4.81 ($\pm$0.84)   & 5.14 ($\pm$0.33)  \\
H$\alpha$ $\lambda$6563      	      	     & 4.07 ($\pm$0.11) & 4.87 
($\pm$0.49)   & 5.19 ($\pm$0.49)   & 6.24 ($\pm$0.64)  \\
$[$S II] $\lambda\lambda$6716, 6731	     & 0.39 ($\pm$0.05) & 0.42 
($\pm$0.03)   & 0.55 ($\pm$0.03)   & 0.71 ($\pm$0.05)  \\
& & & & \\
Flux (H$\beta$)$^{a}$   & 1.65 x 10$^{-13}$ & 1.26 x 10$^{-13}$   & 9.20 x 
10$^{-14}$  & 5.61 x 10$^{-14}$  \\
E(B-V)           & 0.02 ($\pm$0.01) & 0.22 ($\pm$0.02) & 0.20 ($\pm$0.03) &0.24 
($\pm$0.04) \\
\tablenotetext{a}{ergs s$^{-1}$ cm$^{-2}$}
\enddata
\end{deluxetable}

\clearpage
\begin{deluxetable}{lrrrrr}
\tablecolumns{6}
\footnotesize
\tablecaption{Line Ratios from Model Components and Best Fit Composite
(relative to H$\beta$)\label{tbl-3}}
\tablewidth{0pt}
\tablehead{
\colhead{} & \colhead{High Ionization$^{a}$} & \colhead{O$^{++}$$^{b}$} & 
\colhead{Shielded$^{c}$ } & 
\colhead{Composite$^{d}$} &
\colhead{Average Observed$^{e}$}
}
\startdata
CIII $\lambda$977                            & 0.41             & 0.17
              & 0.11                         & 0.22             &  \\
NIII $\lambda$990                            & 0.17             & 0.20
              & 0.05                         & 0.16             & \\
OVI  $\lambda$1036                           &12.81             & 1.63
              & 0.00                         & 4.01             & \\
OV   $\lambda$1216                           & 8.26             & 2.08
              & 0.01                         & 3.11             & \\
Ly$\alpha$ $\lambda$1216                     &33.54             & 4.55 
              &27.59                         &17.56             &14.69  \\
N V $\lambda$1240           	      	     &22.04             & 1.48 
              & 0.02                         & 6.25             & 8.15   \\
Si IV $\lambda$1398                          & 0.03             & 0.02 
              & 0.17                         & 0.05             &incl w/ O IV]\\
O IV] $\lambda$1402                          & 4.53             & 2.39
              & 0.01                         & 2.33             & 1.88    \\
N IV] $\lambda$1486         	      	     & 7.65             & 5.26 
              & 0.46                         & 4.63             & 1.46    \\
C IV $\lambda$1550          	      	     &28.12             & 1.67 
              & 1.38                         & 8.21             & 8.86    \\
He II $\lambda$1640         	      	     & 7.62             & 3.83
              & 0.21                         & 3.85             & 3.48     \\
O III] $\lambda$1663        	      	     & 0.55             & 1.16
              & 0.97                         & 0.95             & 0.38  \\
N III] $\lambda$1750                  	     & 0.66             & 1.58
              & 1.14                         & 1.23             & 0.64   \\
Si III] $\lambda$1892                        & 0.00             & 0.06
              & 0.42                         & 0.14             &incl w/ C III]\\
C III] $\lambda$1909,                        & 1.99             & 1.82 
              & 3.76                         & 2.34             & 4.57   \\
O III] $\lambda$2321                         & 0.04             & 0.09 
              & 0.12                         & 0.09             &incl w/ C II]\\
C II] $\lambda$2326                          & 0.00             & 0.02 
              & 2.17                         & 0.55             & 1.05   \\
$[$Ne IV] $\lambda$2423     	      	     & 0.42             & 0.84 
              & 0.10                         & 0.55             & 2.25   \\
$[$O II] $\lambda$2470      	      	     & 0.00             & 0.01
              & 0.51                         & 0.13             & 0.26    \\
Mg II $\lambda$2800         	      	     & 0.00             & 0.04
              & 1.65                         & 0.43             & 1.25    \\
$[$Mg V] $\lambda$2929         	      	     & 0.03             & 0.02 
              & 0.00                         & 0.02             & 0.14    \\
$[$Ne V] $\lambda$2974                	     & 0.03             & 0.02 
              & 0.00                         & 0.02             & 0.18    \\
He II $\lambda$3204                   	     & 0.43             & 0.24 
              & 0.01                         & 0.23             & 0.30   \\
$[$Ne V] $\lambda$3346      	      	     & 1.10             & 1.06 
              & 0.00                         & 0.81             & 1.24   \\
$[$Ne V] $\lambda$3426      	      	     & 3.00             & 2.90 
              & 0.02                         & 2.20             & 3.28   \\
$[$Fe VII] $\lambda$3588    	      	     & 0.23             & 0.15 
              & 0.00                         & 0.13             & 0.18   \\
$[$O II] $\lambda$3727      	      	     & 0.00             & 0.02 
              & 2.16                         & 0.54             & 0.86    \\
$[$Fe VII] $\lambda$3760    	      	     & 0.32             & 0.19 
              & 0.00                         & 0.18             & 0.27    \\
\tablebreak

$[$Ne III] $\lambda$3869    	      	     & 0.01             & 1.94 
              & 6.81                         & 2.65             & 2.32    \\
$[$Ne III] $\lambda$3967              	     & 0.00             & 0.60 
              & 2.11                         & 0.83             & 0.84    \\
$[$S II] $\lambda$4072                	     & 0.00             & 0.00 
              & 0.48                         & 0.12             & 0.32    \\
H$\delta$ $\lambda$4100 	      	     & 0.26             & 0.25 
              & 0.26                         & 0.26             & 0.31    \\
H$\gamma$ $\lambda$4340      	      	     & 0.47             & 0.46 
              & 0.47                         & 0.47             & 0.53    \\
$[$O III] $\lambda$4363      	      	     & 0.18             & 0.67 
              & 0.55                         & 0.51             & 0.45    \\
He II $\lambda$4686          	      	     & 1.03             & 0.59 
              & 0.03                         & 0.56             & 0.48    \\
H$\beta$                 	      	     & 1.00 	           & 1.00 	
    & 1.00          	 & 1.00    	     & 1.00   \\
$[$O III] $\lambda$5007                      & 4.34              &22.77 
              &29.88                         &19.75              &15.50   \\
$[$N I] $\lambda$5198, 5200     	     & 0.00              & 0.00
              & 2.51                         & 0.63              &  - \\
$[$Fe VII] $\lambda$5721     	      	     & 0.37              & 0.26
              & 0.01                         & 0.22              & 0.35 \\
He I $\lambda$5876           	      	     & 0.00              & 0.06 
              & 0.12                         & 0.06              & 0.16 \\
$[$Fe VII] $\lambda$6087     	      	     & 0.54              & 0.39 
              & 0.01                         & 0.33              & 0.45  \\
$[$O I] $\lambda$6300        	      	     & 0.00              & 0.00 
              & 5.27                         & 1.30              & 0.45   \\
$[$O I] $\lambda$6364                        & 0.00              & 0.00 
              & 1.73                         & 0.43              & 0.22   \\
$[$Fe X] $\lambda$6374                       & 0.90              & 0.08
              & 0.00                         & 0.27              &incl w/ [O I]\\
$[$N II] $\lambda$6548, 6584         	     & 0.00              & 0.11 
              &16.89                         & 4.23              & 4.87  \\
H$\alpha$ $\lambda$6563      	      	     & 2.76              & 2.97 
              & 3.93                         & 3.13              & 5.30   \\
$[$S II] $\lambda\lambda$6716, 6731	     & 0.00              & 0.00 
              & 2.07                         & 0.51              & 0.56    \\
& & & & \\

\tablenotetext{a}{U = 10$^{-1}$, N$_{H}$=5x10$^{4}$, no dust}
\tablenotetext{b}{U = 10$^{-1.3}$, N$_{H}$=1x10$^{5}$, 50\%silicate dust, 75\%
 graphite dust}
\tablenotetext{c}{U = 10$^{-2.35}$, N$_{H}$=5x10$^{4}$, 30\%graphite dust}
\tablenotetext{d}{25\% from hi-ionization, 50\% from O$^{++}$, 25\% from
 shielded}
\tablenotetext{e}{average = 12.5\% each, Nucleus, Position 1; 25\% each,
 Position 2, Position 3}

\enddata
\end{deluxetable}

\clearpage
\plotone{fig1.ps}

\clearpage
\plotone{fig2.eps}

\clearpage
\plotone{fig3.eps}

\clearpage
\plotone{fig4.eps}

\clearpage
\plotone{fig5.eps}

\end{document}